\documentclass[twocolumn,aps,preprintnumbers,amsmath,amssymb]{revtex4}
\usepackage[T1]{fontenc}
\usepackage{graphicx}
\pdfoutput=1
\usepackage{epstopdf}
\usepackage{amsmath}
\usepackage{bm}
\usepackage{color}
\usepackage{float}
\usepackage[caption=false]{subfig}

\usepackage[pdftex,hyperfigures,breaklinks,colorlinks,citecolor=blue,linkcolor=black]{hyperref} % hyperlink

\renewcommand{\figurename}{{\bf Figure}}

\newcommand{\kmax}{k_{\rm core}^{\rm max}}
\begin{document}
%\title{Utilize network analysis to infer language key modules and their circuits from healthy brains' language fMRI}
\title{Core language brain network for fMRI-language task used in clinical applications}

 \author{Qiongge Li$^{1}\footnote{Equal contribution}$, Gino Del Ferraro$^{1,2}\textsuperscript{*}$, Luca Pasquini$^{2,3}$, Kyung K. Peck$^{2,4}$, Hern\'an A. Makse$^{1} \footnote{Equal contribution}$ and Andrei I. Holodny$^{2,5,6}\textsuperscript{$\dagger$}$}
 
\date{\today}

\affiliation{ $^{1}$ Levich Institute and Physics Department, City College of New York, New York, NY 10031}

\affiliation{$^{3}$Department of Radiology, Memorial Sloan Kettering Cancer Center, New York, NY 10065, USA}   

\affiliation{$^{4}$Neuroradiology Unit, NESMOS Department, Sant'Andrea Hospital, La Sapienza University, Rome, RM 00189, Italy}

\affiliation{$^{5}$Department of Medical Physics, Memorial Sloan Kettering Cancer Center, New York, NY 10065, USA}

\affiliation{$^{6}$New York University School of Medicine, New York, NY 10016, USA}
\affiliation{$^{7}$Neuroscience, Weill Medical College of Cornell University, New York, NY 10065, USA}

\begin{abstract}
\begin{center}
\textbf{\abstractname}
\end{center}
Functional magnetic resonance imaging (fMRI) is widely used in clinical applications to highlight brain areas involved in specific cognitive processes. 
Brain impairments, such as tumors, suppress the fMRI activation of the anatomical areas they invade and, thus, brain-damaged functional networks present missing links/areas of activation. The identification of the missing circuitry components is of crucial importance to estimate the damage extent. The study of functional networks associated to clinical tasks but performed by healthy individuals becomes, therefore, of paramount concern. These `healthy' networks can, indeed, be used as control networks for clinical studies. 
In this work we investigate the functional architecture of 20 healthy individuals performing a language task designed for clinical purposes. We unveil a common architecture persistent across all subjects under study, which involves Broca's area, Wernicke's area, the Premotor area, and the pre-Supplementary motor area. We study the connectivity weight of this circuitry by using the  $k$-core centrality measure and we find that three of these areas belong to the most robust structure of the functional language network for the specific task under study. 
Our results provide useful insight for clinical applications on primarily important functional connections which, thus, should be preserved through brain surgery.

\end{abstract}

\maketitle

%%%%%%%%%%%%%%%%%%%%%%

\section{Introduction}

%The human brain performs various cognitive tasks such as language, locomotion and higher cognitive function by integrating different subdivision of the cortex and subcortical structures. Each task involves the neurological activation of several anatomical brain areas which have to collectively cooperate in order to encode external stimuli and produce an appropriate cognitive response (Bertolero, Yeo, \& D?Esposito, 2015; Petersen, S., \& Sporns, O, 2015; Mill, R., Ito, T., \& Cole, M. 2017). The identification of which brain areas are involved in a specific tasks and of the way these areas collectively process information has motived decades of cognitive research aimed to determine the functional specialization of brain regions (Ref.). 

Broca's (BA) and Wernicke's area (WA) have long been recognized as essential language centers. Studies of aphasic patients have shown that damage to BA and WA causes loss of ability to produce speech (expressive aphasia) and difficulty understanding language (receptive aphasia), respectively \cite{dronkers2007paul, wernicke1970aphasic}. Further evidence has shown that other secondary and tertiary anatomical brain areas are also involved in language   \cite{friederici2011brain}, including the pre-Supplementary Motor Area (pre-SMA) \cite{hertrich2016role}, the Premotor Area (preMA) \cite{duffau2003role}, and the Basal Ganglia \cite{booth2007role}. Despite these evidences, a full characterization of the language network is still debated \cite{friederici2017language,fedorenko2009neuroimaging}. 

Functional MRI (fMRI) has been largely used to investigate the blood-oxygen-level dependent (BOLD) activation of the human brain, both for clinical and research purposes. Although it cannot fully resolve the issue of `functional specialization' of brain regions by itself, it sheds light on which regions are engaged in certain cognitive processes. Therefore fMRI allows to constrain hypothesis on the structure of the language network.
	
Language has been investigated using both resting state fMRI (rs-fMRI) and tasked-based fMRI (tb-fMRI). The former studies brain activation of subjects at rest \cite{lee2013resting}, whereas tb-fMRI delineates brain areas functionally involved in the performance of a specific task \cite{bookheimer2002functional}. Task-based fMRI is task-dependent, {\it i.e.} different language tasks may activate different areas involved in language function \cite{xiong2000intersubject}. %This %is one of the reason why a full agreement on the areas involved in language has not been achieved yet (Ref.) and it 
%is not necessarily an innate limitation of the fMRI technique but, rather, a result of the complexity of the human brain that uses several areas for different tasks in a very involved way.  
Consequently, clinical studies employ a specific class of language tasks which has been shown to produce robust activation in individual participants and thus facilitate the localization of the language-sensitive cortex \cite{brennan2007object,ramsey2001combined}. 

In this paper we analyze fMRI scans of 20 healthy individuals who perform the same language task designed for clinical purposes. From the correlation of the BOLD signal we construct the functional connectivity network for each subject, which is standardly employed to investigate statistical interdependencies among brain regions \cite{bullmore2009complex,hermundstad2013structural,gallos2012small}. 

The motivation for this study is to use the resulting functional connectivity of these healthy individuals as a benchmark for clinical study. Brain pathologies indeed (e.g. brain tumors \cite{wang2013group}, strokes \cite{tombari2004longitudinal}, epilepsy \cite{rosenberger2009}) affect functional connectivity by disrupting functional links and reducing the fMRI activation of brain areas (e.g. neurovascular de-coupling effect due to brain tumor \cite{aubert2002modeling}). The reconstruction of the functional connectivity in clinical cases, therefore, is influenced by the presence of the brain pathology \cite{wang2013group}. To better understand which are the functional damages produced by the brain impairment it is important to have, as a benchmark, functional networks of healthy individuals performing the same language task normally used  for clinical cases. The comparison between healthy control and patient's functional network relative to the same task might, among others, guide tumor resection to preserve functional links. 

Motivated by these considerations we investigated which is the language functional architecture shared among healthy subjects, {\it i.e.} the functional subnetwork that persists in each analyzed individual beyond the inter-subject variability. This architecture is indicative of a `core'-structure for the language task under study shared across individuals.

Furthermore, we aim to uncover the functional connectivity of the subdivisions of the Broca's area (pars-opercularis (op-BA) and pars-triangularis (tri-BA), i.e Brodmann area 44 and 45 respectively), which plays a pivotal role in language function \cite{dronkers2007paul,friederici2011brain}. Previous studies based on fMRI showed that BA's subdivisions perform different functions in language processing. Newman {\it et al}. \cite{newman2003differential} showed that tri-BA is more implicated in thematic processing whereas op-BA is more involved in syntactic processing. Studies based on transcranial magnetic stimulation have shown that op-BA is more specialized in phonological tasks and tri-BA more in semantic tasks \cite{devlin2003semantic,gough2005dissociating, nixon2004inferior}. Patients who show speech impairment often have direct damage to the Broca's area. Thus, understanding how BA-subdivisions are functionally wired to other brain regions in healthy controls would help better clarifying the effect of brain pathologies on this decisive language area. 

From our analysis we find that the functional architecture shared by most of the subjects under study wires together Broca's area (op-BA and tri-BA), Wernicke's area, the pre-Supplementary Motor Area, and the Premotor area. By investigating network properties at the subject level we find that, in each individual functional network, these areas belong to an innermost `core', more specifically the maximum `$k$-core' of the functional connectivity, which is a robust and highly connected sub-structure of the functional architecture. The $k$-core measure has received vast attention in network analysis since it provides a topological notion of the structural skeleton of a circuitry \cite{kitsak2010identification,Pittel1996SuddenEO,rubinov2010complex, dorogovtsev2006k}. More recently, the maximum $k$-core has been related to the stability of complex biological systems \cite{morone2019k} and of resilient functional structures in the brain \cite{lucini2019brain}. Our results demonstrate that the functional architecture which persists beyond inter-subject variability is part of the maximum $k$-core structure, an innermost highly connected sub-network, associated with system's resilience and stability \cite{morone2019k}. 

Overall, our findings identify a group of functional regions of interest (fROIs) linked together in a functional circuitry that have a decisive role for the language task used in clinical applications. 
%This network goes beyond inter-subject variability, in truth it is the persistent structure shared across individuals. From a network theory perspective and dynamical system theory, it is the most resilient and robust sub-structure and, thus, clinical procedures should strive to preserve it.    

%%%%%%%%%%%%%%%%%%%
	
\section{Materials and Methods}\label{sec:data}
		
The study was approved by Institutional Review Board and an informed consent was obtained from each subject.
The study was carried out according to the declaration of Helsinki. Twenty healthy right-handed adult subjects (13 males and 7 females; age range 36 years, mean = 36.6; SD = 11.56) without any neurological history were included.

%%%%%%%%%%%%%%%%%
	
\subsection{Functional task}

\begin{figure}
\includegraphics[width=0.45\textwidth]{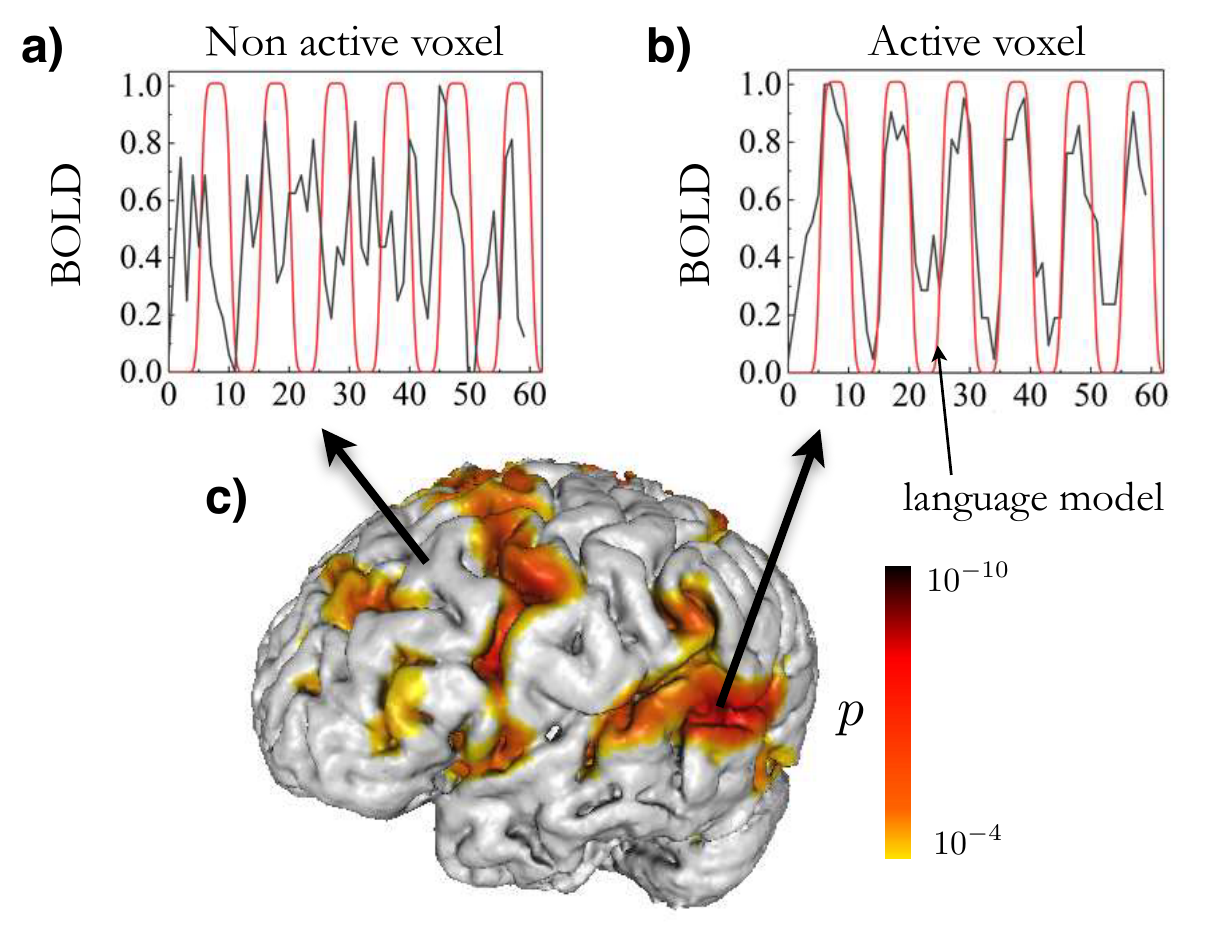}
\caption{{\bf Activation map for a representative subject}. BOLD signal for a non-active and active voxel are shown respectively in panel a) and b) together with the smoothed boxcar language model which depicts the auditory stimulus. c) 3D visualization of the brain with fMRI active areas and corresponding p-values.}
\label{fig:act_map}
\end{figure}

For the fMRI task, all subjects performed a verbal fluency task using verb generation in response to auditory nouns. During the verb generation task, subjects were presented with a noun (for example, `baby') by oral instruction and then asked to generate action words (for example, `cry', `crawl', etc...) associated with the noun. Four nouns were displayed over six stimulation epochs with each epoch lasting 20 seconds, which allowed for a total of 24 distinct nouns to be read over the entire duration. Each epoch consisted of a resting period and a task period (see BOLD activation in Fig. \ref{fig:act_map}, panel a-b). In order to avoid artifacts from jaw movements, subjects were asked to silently generate the words. Brain activity and head motion were monitored using Brainwave software (GE, Brainwave RT, Medical Numerics, Germantown, MD) allowing real-time observation.

% FIGURE %%%%%%%%%%%%

%%%%%%%%%%%%%%%%
	
\subsection{Data acquisition}

A GE 3 T scanner (General Electric, Milwaukee, Wisconsin, USA) and a standard quadrature head coil was employed to acquire the MR images. Functional images covering whole brain were acquired using a T2* weighted gradient echo echo-planar imaging sequence (repetition time (TR)/echo time (TE) $= 4000/40$ ms; slice thickness $= 4.5$ mm; matrix $= 128 \times 128$; FOV $= 240$mm). Functional matching axial T1-weighted images (TR/TE $= 600/8$ ms; slice thickness = $4.5$ mm) were acquired for anatomical co-registration purposes. Additionally, 3D T1-weighted SPGR (spoiled gradient recalled) sequence (TR/TE = $22/4$ ms; slice thickness $= 1.5$ mm; matrix $= 256 \times 256$) covering entire brain were acquired. 
 
 %%%%%%%%%%%%%
 
\subsection{Data processing}\label{sec:}
 
Functional MRI data were processed and analyzed using the software program Analysis of Functional NeuroImages (AFNI) \cite{cox1996afni}. Head motion correction was performed using 3D rigid-body registration. Spatial smoothing was applied to improve the signal-to-noise ratio using a Gaussian filter with 4 mm full width of half maximum. Corrections for linear trend and high frequency noise were also applied. To obtain the activation map, BOLD signal changes over time were cross-correlated with a baseline smoothed box-car model representative of the word-generation epochs and period of rests (see Figure \ref{fig:act_map}, panel a-b). Functional activation maps were generated in the individual native space at a threshold of $p< 0.0001$ (see Figure \ref{fig:act_map}, panel c, for a representative subject). %To transfer individual brain coordinates into a standard template space, each dataset was registered to MNI 152 (the Montreal Neurological Institute template 152). 

%%%%%%%%%%%

\subsection{Network construction}\label{sec:}
	
The following sections describe the functional network construction. In Sec. \ref{sec:individual} we first describe how to create, from the fMRI signal of the active voxels, a brain network for each individual separately. Section \ref{sec:common} discusses the group analysis or how we obtain, from the individual brain networks, a common architecture which unveils a persistent circuitry across all the single-subject brain networks.

% FIGURE %%%%%%%%%%%%
\begin{figure}[!th]
\includegraphics[width=0.35\textwidth]{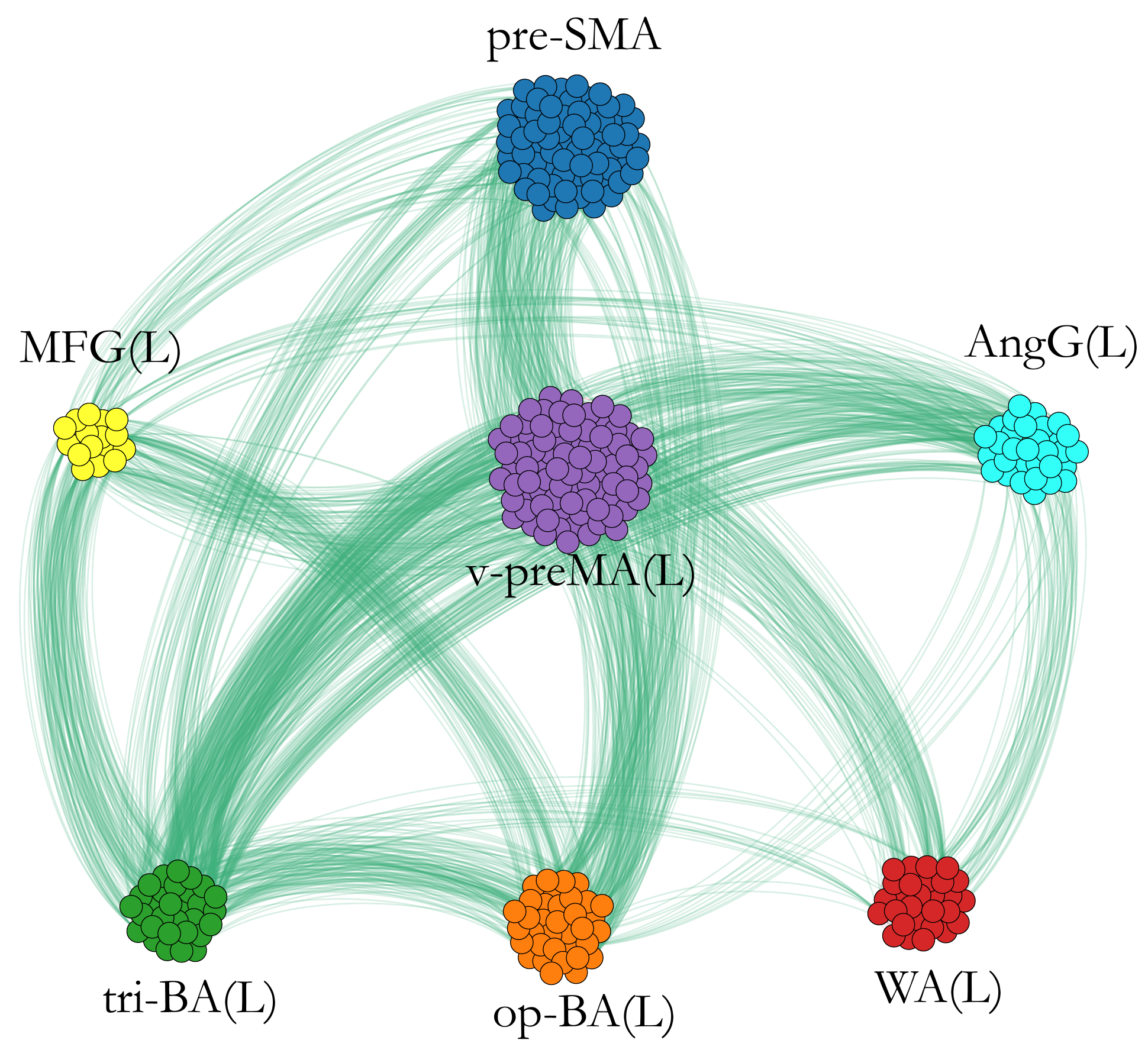}
\includegraphics[width=0.4\textwidth]{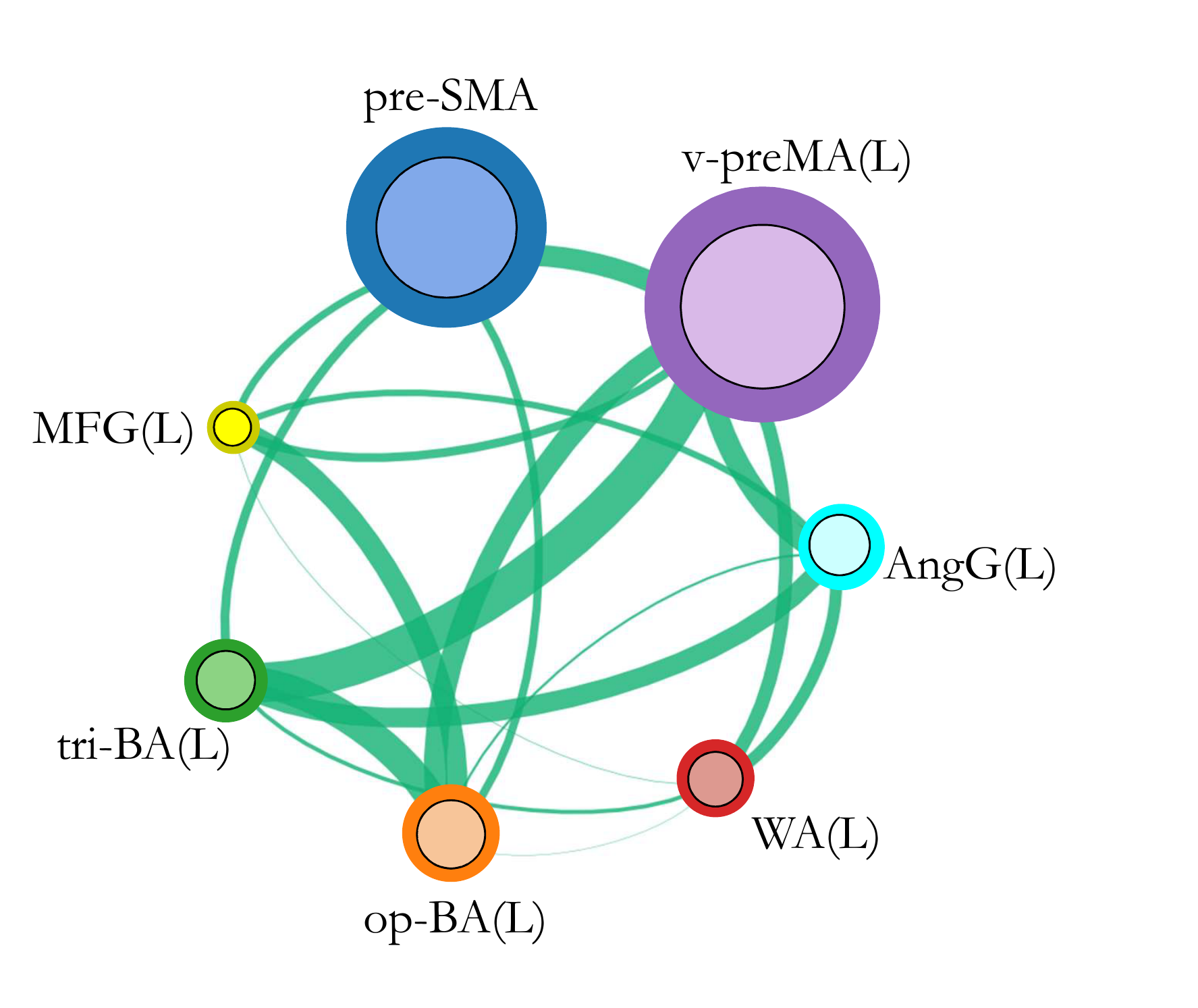}
\caption{{\bf Two visualizations of the individual functional network}. Upper panel: voxel-level network. Each node in the network represents a voxel, each link connects a pair of voxels in different brain modules and it is indicative of functional interdependecy. Links connecting voxels within the same brain module are not visible but exist. Lower panel: fROI-level network for the same voxel-level architecture shown in the upper panel. Voxels belonging to the same anatomical region are grouped into a fROI, represented as a node in the network. Node's size is proportional to the number of voxels in the fROI. Each link's thickness connecting two fROIs is proportional to the sum of links' weight connecting all the voxels in the two fROI (exact definition given in Eq. \eqref{eq:W}).}
\label{fig:representative}
\end{figure}
	
%%%%%%%%%%%%%%%
	
\subsection{Individual brain network construction }\label{sec:individual}

For each subject we construct a functional network, this network can be seen at two different scales or levels: {\it i)} at the {\it voxel-level} and {\it ii)} at the {\it fROI-level}, as we explain in more details in the following. 

At the {\bf voxel-level}, active voxels in the individual activation map ($p<0.0001$) define the nodes of our functional network. Functional links are inferred by thresholding pair-wise correlations between pair of voxels, as standard in the literature \cite{bullmore2009complex,hermundstad2013structural,gallos2012small}. Accordingly, pair of voxels with correlation above a fixed threshold are connected by a link, whose weight is given by the correlation strength \cite{bullmore2009complex,lucini2019brain,del2018finding}. Nearby active voxels are grouped together based on the subject individual anatomy and considered part of the same fROI. Figure \ref{fig:representative}, upper panel, shows a realization of the voxel-level functional network for a representative subject, where voxels part of the same fROI are colored equally. 

We define fROIs within each subject individually, based on the activation and anatomy of the specific subject \cite{lucini2019brain,del2018finding,fedorenko2010new}. For instance, all the active voxel in Brodmann area 22 of the superior temporal lobe define the Wernicke's area fROI. The reason for choosing individual-based fROIs is that group-based ROI-level analysis suffer of inter-subject variability in the location of activation, in contrast, individual subject fROIs analysis can reveal greater functional specificity \cite{fedorenko2010new}. 

At the {\bf fROI-level} a node represents an entire fROI, {\it i.e.} a group of nearby active voxel in the same anatomical area. At this level, a functional link connects two fROIs iff there exist at least a link, at the voxel-level, between a pair of voxels in the two fROIs.
The functional link's weight between two fROIs $i$ and $j$ ($W_{ij})$ is defined as the sum of all the functional links' weight ($w_{lm}$) connecting all pair of voxels ($l,m$) between the two fROIs, normalized by the sum of the two fROIs' size:
\begin{equation}
W_{ij} = \frac{\sum_{l,m \in \{i \leftrightarrow j\}} w_{l,m}}{\rm size(fROI_{\it i}) + size(fROI_{\it j})}
\label{eq:W}
\end{equation}
For each individual we then normalize each $W_{i j}$ by the value of the largest $W$ for that individual ($W^{\rm max}$):
\begin{equation}
 \tilde{W}_{ij} = \frac{W_{ij}}{W^{\rm max}}, \qquad \textrm{for all {\it i} and {\it j} fROIs}
 \label{eq:Wrescale}
\end{equation}
In this way the link's weight scale is the same across subjects (see Supplementary Table \ref{table:20individuals}) and the maximum weight is $\tilde{W} = 1$ in each individual. Figure \ref{fig:representative}, lower panel, illustrates the functional network at the fROIs-level for a representative subject. 

For each individual, we are interested in uncovering the functional architecture of the subdivision of the Broca's area, i.e tri-BA and op-BA, which correspond to all active voxels in Brodmann area 44 and 45, respectively. Each of these sub-area has been associated with different language processes in previous studies \cite{newman2003differential,devlin2003semantic,gough2005dissociating, nixon2004inferior}. Through our analysis we aim to find out the specificity of their functional connectivity, to unveil whether their different engagement in language processing may be associated to a different functional wiring with the rest of the brain. Thus, when building the individual functional network, we group the active voxels of the BA into two different and separate fROIs: op-BA and tri-BA (see Fig. \ref{fig:representative}).

We named the fROI according to their main anatomical boundaries as follows. 
%Each active area was consistent across subjects in term of location. 
We retained the classical designations
of BA (Brodmann area 44-45, inferior frontal gyrus) and WA (Brodmann area 22, superior temporal gyrus) as these designations still predominate in neurosurgery, which dominates clinical practice \cite{friederici2011brain}. We defined the ventral Premotor area (v-preMA) as the ventral portion of the premotor cortex, which includes the inferior part of Brodmann's area 6, centered on the posterior-most portion of the middle frontal gyrus (MFG) \cite{friederici2011brain}. The superior portion of Brodmann's area 6 was considered dorsal premotor area (d-preMA). The anterior-most part of the middle frontal gyrus was identified as anterior middle frontal gyrus (aMFG). The pre-SMA was defined within the medial frontal cortex, at the level of Brodmann area's 6 \cite{nachev2008functional}. The precentral gyrus was identified with Brodmann's area 4, the supramarginal gyrus was identified with Brodmann's area 40 and angular gyrus with Brodmann's area 39  \cite{friederici2011brain}.
The deep opercular cortex (DOC) included the innermost portion of the frontal operculum \cite{friederici2011brain}. 
Active areas that support non-linguistic processing, as the visual and the auditory cortex, were excluded from the analysis \cite{fedorenko2010new,fedorenko2009neuroimaging}. These areas are indeed activated because the subject is presented with auditory stimuli and may keep the eyes open. 

The same functional network construction as described above is carried over for all the 20 subjects individually, both at the voxel- and fROI-level. Next, we carry a group analysis to identify the common functional network shared across individuals, beyond the inter-subject variability, as described in following section.

%%%%%%%%%%%

\subsection{Common network construction across subjects }\label{sec:common}

Our interest in studying functional networks for single individuals performing language tasks is aimed to uncover functional architectures which are persistent across healthy subjects and could be useful and informative when dealing with clinical cases. Individual functional networks have innate subject variability (e.g. some subject activates one specific area or has a functional link while another does not). Therefore, after we reconstructed the individual functional networks, we performed a group analysis at the fROI-level by investigating which is the persistent set of links and brain areas across subjects or, in other words, which functional sub-architecture is common among all the individuals.  

This functional architecture is informative of which areas and functional links persist beyond the inter-subject variability and therefore it represents a language `core' structure for the specific language task under study. Accordingly, surgical intervention, as for instance tumor resection, should operate by preserving such `core' structure existing across healthy controls. In addition, functional damages to this structure due to brain pathologies - and observed from the functional connectivity of the patient - may be informative of the damage extent (e.g. a missing functional link in the `core' may signify a larger harm than a missing connection between more peripheral areas not in the `core'). 

We name this most persistent functional architecture across subjects, at the fROI-level, `{\it common network}'. This common architecture is defined retaining a pair of fROIs and a functional link connecting them only if these areas and link are present across subjects.

The weight of the functional link connecting two fROIs $i$ and $j$ in the common network ($W_{ij}^{\rm C}$) is defined as the average of the $\tilde{W}_{ij}$ connecting those fROIs across subjects:
\begin{equation}
W_{ij}^{\rm C} = \frac 1 N \sum_{l=1}^N \tilde{W}_{ij}^{(l)}
\end{equation}
where $N$ is the total number of individuals. We report and discuss the results of this quantitative analysis in the following sections.

%%%%%%%%%%

\section{Results}\label{sec:results_individuals}
\subsection{Individual networks}

For each individual we observe fMRI activation in both hemispheres, however, left dominance is clearly observed, as expected since all the subjects are right handed \cite{isaacs2006degree,knecht2000handedness}. The number of left hemisphere areas of activation regions is greater and in most cases their frequency of activation is greater as well. 
%Deep Opercular cortex - Insulo Opercular Angle -Deep frontal opercular

Active fROIs across subjects include, in alphabetic order: Angular Gyrus (L), Broca's Area (L), (op-BA and tri-BA), Broca's Area (R), Caudate (L and R), Deep Opercular Cortex (L and R) \cite{friederici2011brain}, aMFG (L and R), pre-Central Gyrus (L and R), ventral and dorsal preMA (L), ventral preMA (R), pre-SMA, Supra Marginal Gyrus (L and R), Wernicke's Area (L and R). Detailed information on the frequency of activation of each area across subjects is summarized in Supplementary Table \ref{table:areas}. In the following, for brevity, we will refer to left hemisphere brain areas simply with the name of the areas, omitting the specification (L). 

The functional network for a representative subject at both the voxel- and the fROIs-level is shown in Fig. \ref{fig:representative}. All the single subjects functional connectivity at the fROI-level for each of the 20 healthy individuals considered in our study are shown in Supplementary Figure \ref{fig:20individuals} and all the connectivity values between pairs of fROIs are reported in Supplementary Table \ref{table:20individuals}. 

We observe that, overall, the preMA  is the most connected area across subjects, in terms of connectivity weight. In 8 over 20 individuals the strongest functional connection is between preMA - op-BA and in 7 over 20 cases is between preMA - pre-SMA. In total, the preMA turns out to have the strongest connection with some other area in 17 out of 20 subjects, in 3 cases the strongest functional connection is between op-BA and tri-BA.

Wernicke's area is known to structurally connect to BA through the arcuate fasciculus, a bundle of axons linking the inferior frontal gyrus with the superior temporal gyrus. We investigated the functional connections of the BA-subdivisions with the rest of the brain and, with focus on WA, we find that op-BA connects to WA in 18 out of 20 subjects, while tri-BA connects to WA in 15 out of 20 individuals. In terms of connectivity weight, in 10 out of 20 subjects WA connects more strongly with op-BA than to tri-BA, whereas in 7 subjects we have the opposite finding, tri-BA connects more to WA than the opercular counterpart. In 2 individuals the functional connectivity of op-BA and tri-BA to WA is, instead, approximately the same. One subject does not show WA activation at all.

Regarding other relevant areas as preMA and pre-SMA we find that the connectivity frequency of these areas with op-BA and tri-BA is about the same. Indeed the preMA connects to op-BA in 18 subjects and to tri-BA in 17 out of 20. The pre-SMA connects to op-BA in 19 subjects and to tri-BA in 18 individuals. So, overall, the connectivity frequency of the BA-subdivisions with preMA and pre-SMA is similar. In terms of connectivity weight, op-BA connects more strongly with both preMA and pre-SMA compared to tri-BA. Thus, although the BA-subdivisions connect to preMA and pre-SMA with about the same frequency across subjects, op-BA has - overall - a larger connectivity weight. 

% FIGURE %%%%%%%%%%%%
\begin{figure}[!t]
\centering
\includegraphics[width=0.5\textwidth]{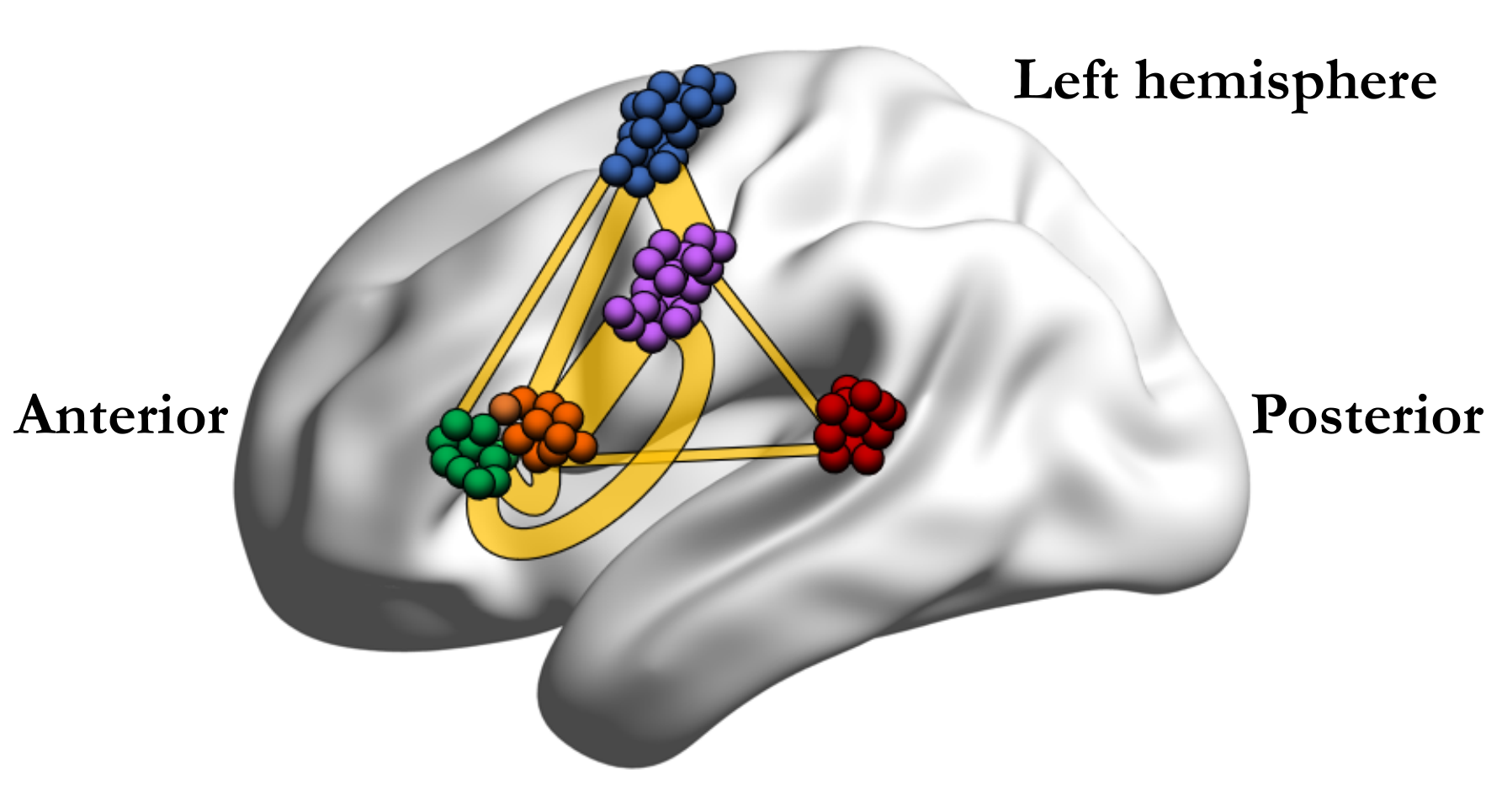}
\includegraphics[width=0.5\textwidth]{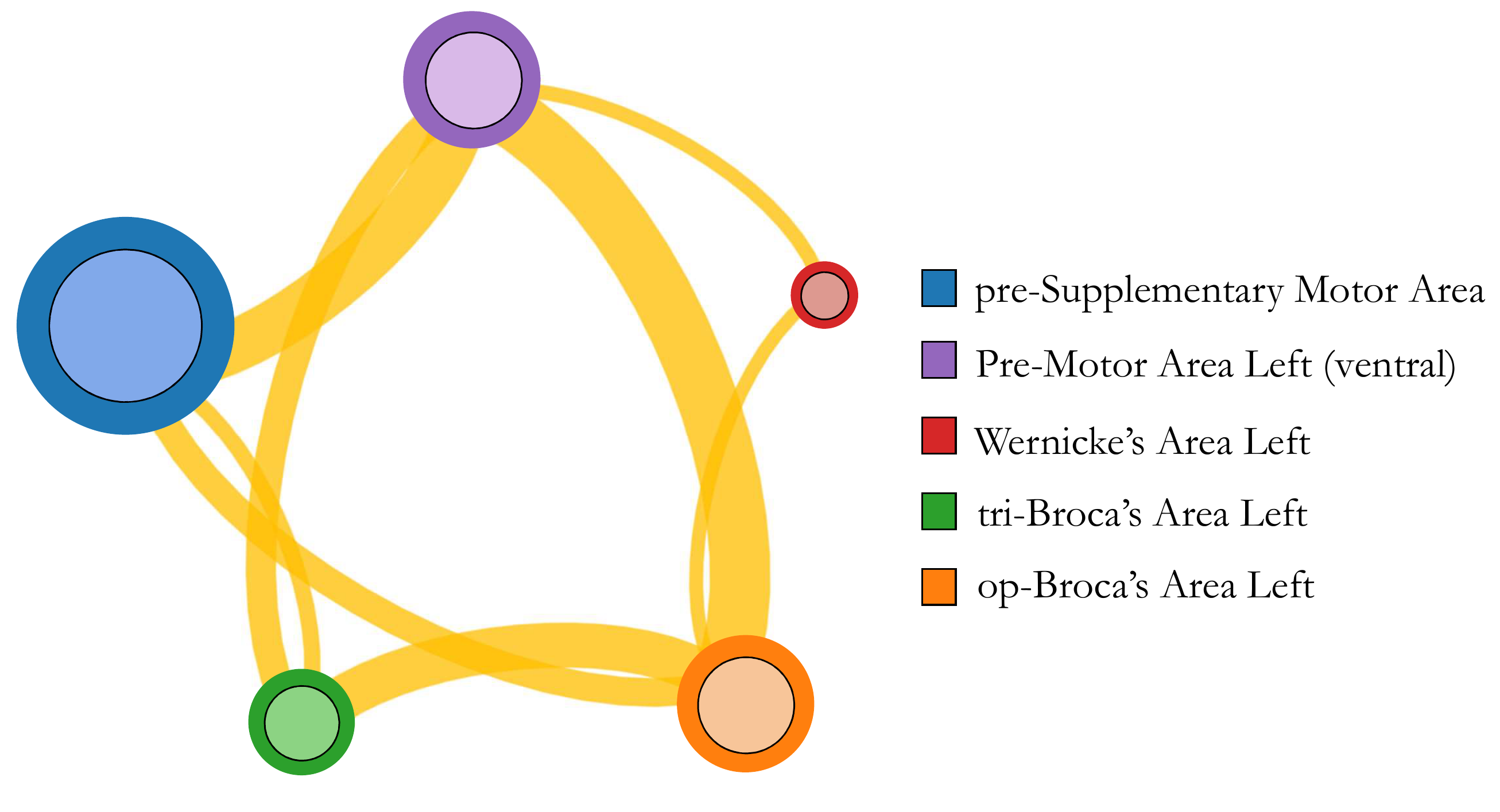}
%\vspace{-1.5cm}
\caption{{\bf Common network across subjects for the language task under study}. The figure illustrates the functional network, beyond inter-subject variability, shared across individuals (seventeen out of twenty). The weight of a link connecting two fROIs is proportional to the average of the functional links connecting those fROIs across subjects. Upper panel: fROIs are located on their anatomical location on the brain. Lower panel: pictorial illustration of the network in the upper panel, with the fROIs equally spaced on a plane.}
\label{fig:common3D}
\end{figure}

%%%%%%%%%%%%%%%%%%

\subsection{Common network across subjects and functional subdivisions of Broca's area}\label{sec:}
	
The common network at the fROI-level, as described in Sec. \ref{sec:common}, is made by those fROIs and links  present (persistent) across the majority of subjects.  As a result of the left-dominance at the individual level, no consistent overlap of right-hemisphere activation has been found across subjects. 

We find that the persistent structure across individuals (seventeen over twenty), beyond inter-subject variability, is made by op-BA, tri-BA, WA, preMA, and pre-SMA connected together in a functional architecture (see Fig. \ref{fig:common3D}). This circuitry represents the `core'-structure for the specific clinical language task under investigation since it is the functional architecture that prevails in nearly all subjects. We find this network in seventeen over twenty subjects and not in all of them because three subjects show lack of activation for either the op-BA (1 case), the tri-BA (1 case), or neither WA nor tri-BA (1 case). The common network shown in Fig. \ref{fig:common3D} is therefore the one prevailing in closely all the subjects and, thus, the functional structure that is persistent beyond inter-subject variability. This conclusion is additionally supported by further findings we obtained on a study conducted on bilingual healthy subjects when they speak their native language \cite{LiBilingual}. 

In terms of functional connectivity, the strongest connectivity weight in the common network ($W^C_{\rm max}$) is between op-BA and preMA ($W^C_{\rm max} = 0.74 \pm 0.31$, where the average is made across all the subjects that have such link). The triangular BA also connects with the preMA but with about half of the magnitude ($W^C = 0.37 \pm 0.29$). Detailed information on the functional connection of the other areas in the common network is reported in Supplementary Table \ref{table:common}. Broca's area has been longly recognized as a central language area, its strong connectivity with the preMA(L) is of particular interest since the preMA(L) has been more recently identified as an area with dominant role for language \cite{duffau2003role}. We discuss this result further in Section \ref{sec:dis_common}. 

When we look at the connectivity of the BA-subdivisions with Wernicke's areas, a primary area for language comprehension, we observe that, in the common network, WA only connects to op-BA. This reveals the existence of larger co-activation of the BOLD signal between these two areas that might also be driven by their spatial vicinity (WA is anatomically  closer to op-BA than to tri-BA). More detailed, as discussed in Sec. \ref{sec:results_individuals}, at the individual level we find that WA connects to tri-BA in $15$ out of $20$ cases whereas it connects to op-BA in $18$ out of $20$ cases. Therefore, overall, BA-subdivisions both connect to WA in several different individuals with a slightly larger presence of WA - op-BA connectivity across subjects.  In terms of connectivity weight, when we count only subjects where both op-BA and tri-BA connect to WA we find that op-BA connects slightly more strongly to WA compared to tri-BA ($W^C = 0.17 \pm 0.23$ {\it vs} $W^C = 0.15 \pm 0.20$, respectively).

Furthermore, we observe that op-BA has a larger connectivity than tri-BA both on the number of connections with the rest of the areas in this network (4 {\it vs} 3 respectively, the extra one being WA - op-BA) and in terms of functional connectivity weight. Indeed, the average connectivity of the op-BA, across subjects and across areas, in the common network is $W^C =0.45 \pm 0.25$ ($W^C = 0.55 \pm 0.20 $ without the link WA - op-BA) whereas the comprehensive connectivity weight of the tri-BA is $W^C =0.32 \pm 0.18$.

%and this fact prevents the possibility of having the above structure persistent across all the cases under study. When all 20 patients are considered, we find that the common network is made by BA, PreM, and SMA functionally wired together. Wernicke's area is one of the most important regions for language production (Ref.) and is very often found active in fMRI analysis. We interpret the missing activation of this area in one over twenty subjects as an experimental anomaly and, therefore, hereafter we consider the common network (the `core' structure) made by the BA, WA, PreM, and SMA as we find in nineteen over twenty cases (see Fig. \ref{}). 

\begin{figure}
\includegraphics[width=0.37\textwidth]{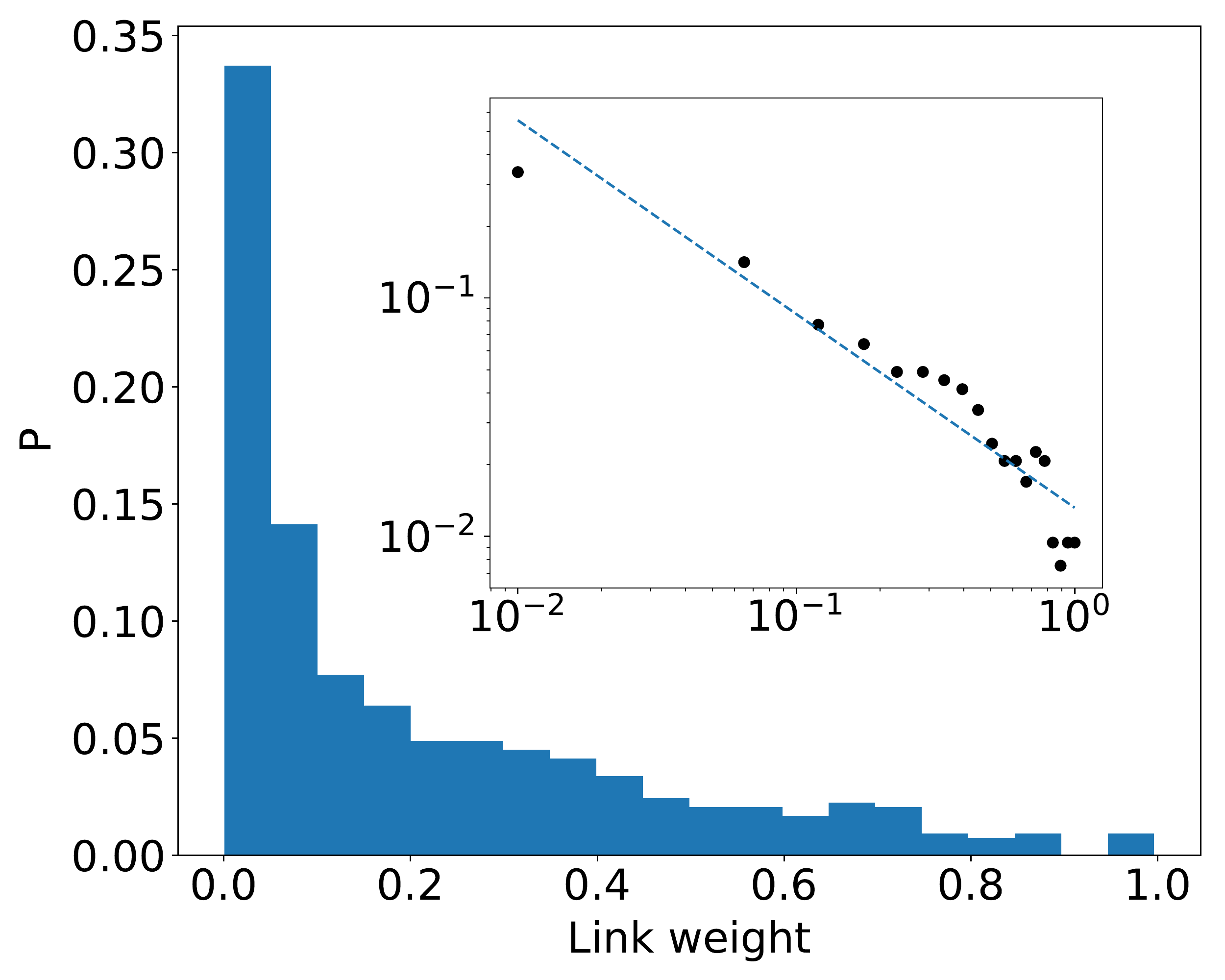}
\caption{{\bf Functional connectivity weight distribution for all the subjects.} Distribution of the functional link's weight across all  subjects. The distribution is long tailed and this explains the large variance values in Supplementary Table \ref{table:common}. Inset: same distribution plotted in a log-log scale, data aligns on a straight line which indicates a power low behaviour for such distribution.}
\label{fig:weights}
\end{figure}

Finally, we observe that the average values for the common-network functional weights reported in Supplementary Table \ref{table:common} have large standard deviations (magnitude comparable with the mean). This result signals a large inter-subject variability for the weight of the single functional link across subjects. To investigate this further we plot the empirical distribution of all the functional links' weights across subjects and observe that it displays a long tail shape (see Fig. \ref{fig:weights}), which is explicative of the large standard deviation values. Since all the individual links' weight are normalized to one, we plot the distribution in Fig. \ref{fig:weights} up to the value one excluded, because otherwise this distribution would show a second pick in one due only to the normalization procedure.

% FIGURE  %%%%%%%%%%%%

\subsection{The common network is part of the maximum $k$-core: the most resilient architecture}

The notion of $k$-core in theoretical physics has been used as a fundamental measure of centrality and robustness within a network \cite{morone2019k}. Since it was firstly introduced in social sciences \cite{seidman1983network} has been used in several contexts \cite{kitsak2010identification}, as in random network theory \cite{Pittel1996SuddenEO} or to describe large-scale structure in the brain \cite{hagmann2008mapping}. 

\begin{figure}
\includegraphics[width=0.4\textwidth]{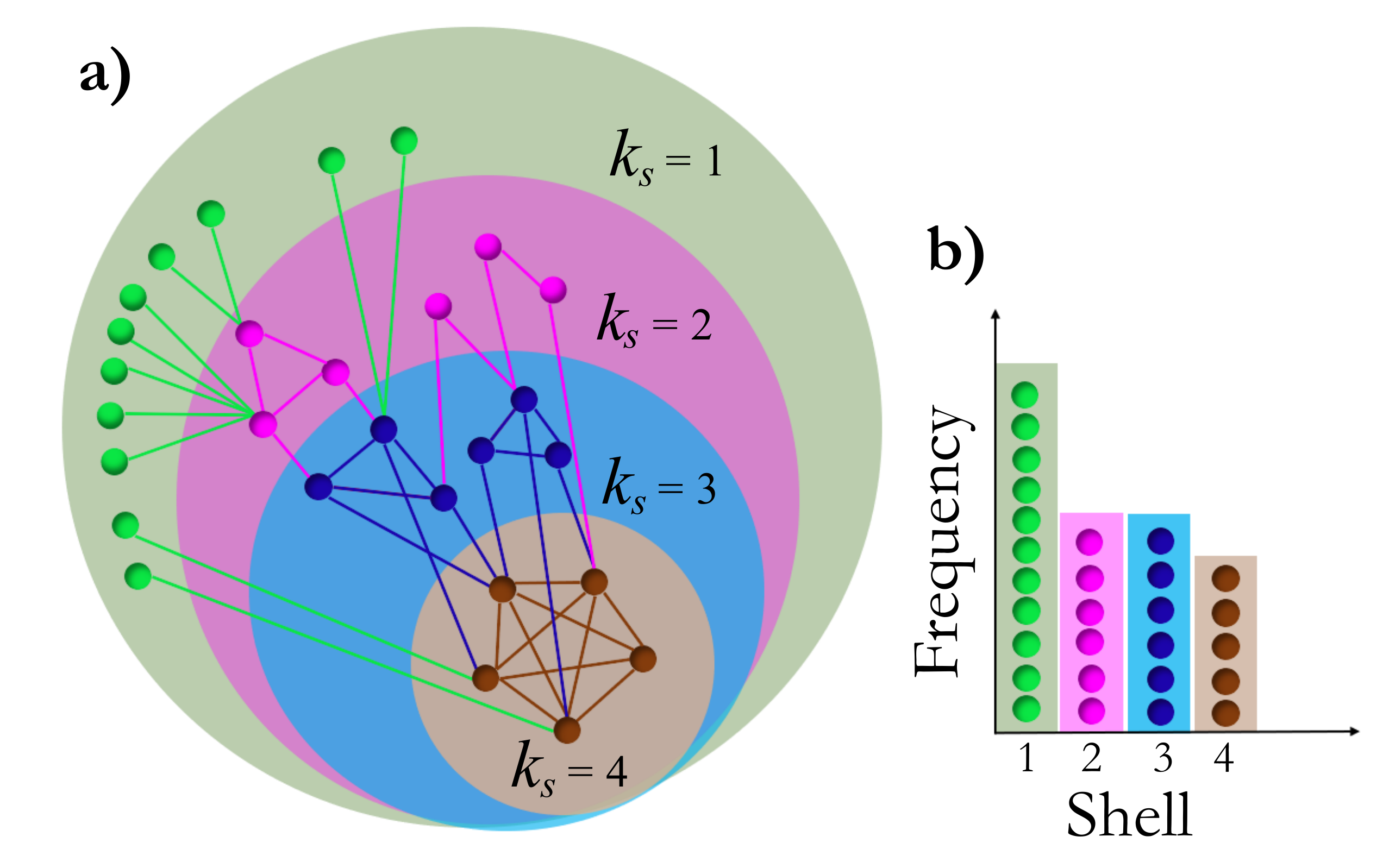}
\caption{{\bf $k$-core and $k$-shell of a network}. Panel a) illustrates pictorially a network.  Nodes in the same disks have the same $k$-core. A $k$-core structure includes sub-networks with higher $k$'s, so the 1-core includes the 2-core which, in turn, includes the 3-core and so forth. Nodes which are in the $k$-core but not in the $k+1$-core are called $k$-shell and are colored differently. The maximum $k$-core coincides with the maximum $k$-shell, in this network is $\kmax=4$ and depicted with brown nodes. Panel b) illustrates pictorially the construction of the $k$-core histogram shown in Fig. \ref{fig:kcorehistogram}. Note that here nodes in each $k$-shell are colored differently, whereas in Fig. \ref{fig:kcorehistogram} different colors indicate nodes in different fROIs, piled up according to their $k$-shell as in this panel.}
\label{fig:kcore}
\end{figure}

The $k$-core of a given architecture is defined as the maximal sub-graph, not necessarily globally connected, made of all nodes having degree (number of connections) at least $k$. In practice, the $k$-core sub-graph can be derived by removing from the network all nodes with degree less than $k$. The removal of these nodes reduces the degree of their neighbours and if the degree of the latter drops below $k$ then also these nodes should in turn be removed. The procedure iterates until there are no further nodes that can be pulled out from the network. The remaining graph is the $k$-core of the network. A $k$-core structure includes sub-networks with higher $k$'s, {\it i.e.} $k + 1$, $k+2$, etc... For instance, the 1-core includes the 2-core which, in turn, includes the 3-core and so forth (see Fig. \ref{fig:kcore}). In each $k$-core, nodes in the periphery (not included in the $k+1$-core) are called $k$-shell ($k_s$). Thus, in each network, $k$-core (and $k$-shell) structures are nested within each other with increasing $k$. The innermost structure of the network corresponds to the graph with the maximum $k$-core that is also a topological invariant of the network \cite{dorogovtsev2006k}. Figure \ref{fig:kcore}, panel a, illustrates $k$-core and $k$-shell structures in a simple explanatory network.

Recently, the maximum $k$-core ($k_{\rm core}^{\rm max}$) has been linked to the most resilient structure of biological systems with positive interactions \cite{morone2019k} and, in an fMRI study of human brains, the $\kmax$ of the functional connectivity for a visual-task based experiment has been found to be the most robust structure, which remains active even during subliminal conscious states (subject not aware of seeing images) \cite{lucini2019brain}. 

% FIGURE  %%%%%%%%%%%%

\begin{figure}
\includegraphics[width=0.4\textwidth]{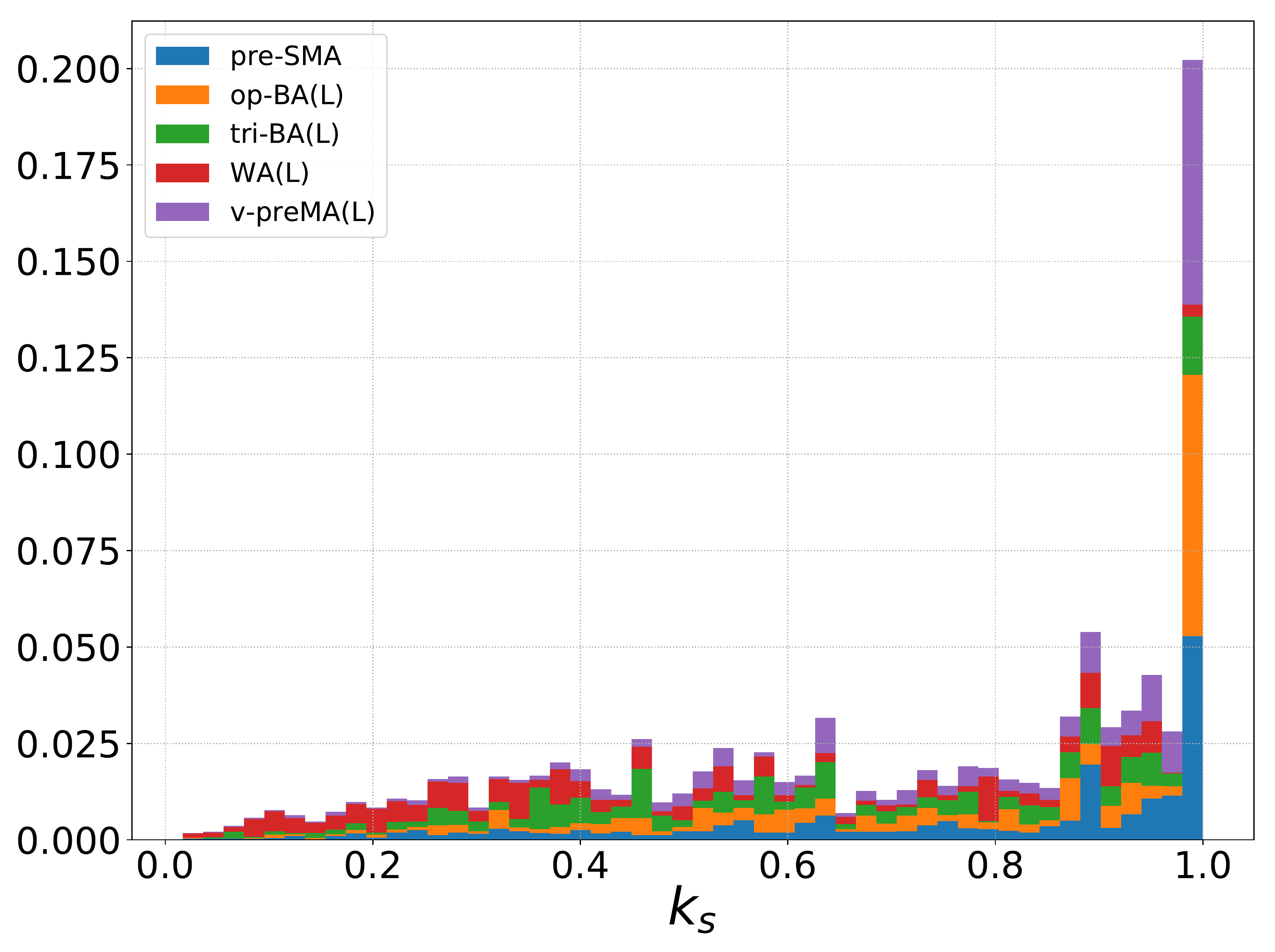}
\caption{{\bf k-shell occupancy}. The histogram shows the k-shell occupancy for nodes in the four fROIs of the common network of Fig. \ref{fig:common3D}. Overall, the majority of the nodes of this structure are located in the maximum $k$-shell, which coincides with $\kmax$, a quantity linked to the robustness of a complex network \cite{morone2019k}. Of the four fROIs of the common network, the pre-SMA, op-BA, tri-BA, and ventral preMA are mostly part of the $\kmax$. Wernicke's area (WA) is more an outlier, it is mostly located in lower $k$-shells and minimally located in the $\kmax$.}
\label{fig:kcorehistogram}
\end{figure}

Motivated by these recent findings we pruned each voxel-level individual functional network till the maximum $k$-core structure and we investigated to which $k$-core each node (voxel) belongs to. We focused on the areas part of the common network (BA, WA, pre-SMA, and preMA) because these are the fROIs which form a persistent language structure across individuals, as shown in Fig. \ref{fig:common3D}. We aim to explore whether these regions are part of some significative $k$-core structure which might shed light on the architecture of the network. Our goal is to investigate, across subjects, which fROIs characterizes the occupancy of each $k$-shell and, thus, we proceed as follows. For each individual network we compute the $k$-core and $k$-shell of all the nodes (voxels) as described above. Each subject has, in general, a different $\kmax$ ($k$-shell) thus, to homogenize the $k$-cores across subjects, we divide each $k$-core by the individual $\kmax$. In this way, the $k$-core ($k$-shell) value of each individual goes from $0$ to $1$. In Fig. \ref{fig:kcorehistogram} we then plot the total $k$-shell occupancy for all the individuals and we color differently the contribution of each fROIs, in order to visualize to which $k$-shell they belong to. 

Results in Fig. \ref{fig:kcorehistogram} show that the maximum $k$-shell (which is in turn the maximum $k$-core) is the most populated of all the $k$-shells of the common network. More importantly, if we look at each area individually, we observe that the largest concentration of pre-SMA, op-BA, tri-BA, and v-preMA is in the maximum $k$-shell. Among the areas of the common network, WA is the only area that does not appear in the $\kmax$ but, rather, populates smaller $k$-shell values. 

In reference \cite{morone2019k} the authors have shown that, for complex networks with positive couplings, the $\kmax$ of the network is the most resilient structure under decreasing of the coupling weight. In our functional networks, all the links are obtained through thresholding of pair-wise correlations which, from our findings, turn out to be all positive. This is due to the fact that the BOLD signal is extracted from a task-based fMRI experiment, stimulated by an external input. In this way, active voxels are those mostly correlated with the task model and, when computing pair-wise correlations among voxels correlated with the same external stimulus, most likely one finds positive correlations, as we observe from our data analysis. This result allows us to interpret the functional networks wired by positive interactions and, therefore, the theory of \cite{morone2019k} applies. Accordingly, we can interpret the maximum $k$-core structure of our network as the most resilient one under decreasing of the correlation weight.

In other words, the circuitry made by the pre-SMA, BA, and the preMA represents the most robust structure of the functional network. Wernicke's area, although it is part of the common network, it does not lay in the $\kmax$ of the network, probably due to its more peripheral anatomical location compared to the other fROIs of this common architecture. Therefore, although it is one of the most important areas for language, it is not part of the most resilient `core'.

% FIGURE  %%%%%%%%%%%%

%%%%%%%%%%%%%%%%%

\section{Discussion}\label{sec:}

In this study, we reconstructed the functional language network of 20 healthy subjects from tb-fMRI data providing information about the functional connectivity between active areas on fMRI maps, with both a voxel-  and a fROI-level resolution. The language task designed for the experiment is customarily used in clinical cases and has shown to produce robust activation in previous studies \cite{brennan2007object,ramsey2001combined,xiong2000intersubject,Dong2019}. Functional activation is generally sensitive to the fMRI task employed, our interest in reconstructing functional networks for this specific task aimed to create benchmark results for healthy individuals which can be used as reference for functional networks effected by brain pathologies. Indeed, brain impairments are known to create damages on the functional connectivity. It is therefore paramount to have healthy functional architectures relative to clinical language tasks in order to make the comparison between healthy and patient's functional networks possible. 

Our main finding is the existence of a common persistent functional network across subjects which wires together BA, WA, ventral preMA, and pre-SMA in the left dominant hemisphere for 17 out of 20 right-handed healthy subjects (see Fig. \ref{fig:common3D}). We interpret this circuitry as a `core' structure for the language task under study since this network persists across nearly all individuals. 

Furthermore, we compute the $k$-core of each node (voxel) in the common network - the maximum value of which has been recently linked to network resilience in ecosystems and fMRI studies \cite{morone2019k,lucini2019brain}. We find that 3 out of 4 areas of the common architecture (specifically pre-SMA, BA, and preMA) are mostly concentrated in the maximum $k$-core of the network (see Fig. \ref{fig:kcorehistogram}). This led us to conclude, following the fundings of Ref. \cite{morone2019k,lucini2019brain}, that these areas are the most robust of the language network in terms of fMRI correlated signal. 

Wernicke's area is a crucial language area and indeed appears as part of the common network across individuals, yet its type of connectivity with the rest of the fROIs in this architecture is slightly different from the connectivity of the other areas. Indeed, overall, WA shares only two connections with other areas in this network, one with BA and one with the preMA, whereas each of the other fROIs has at least 3 total functional connections. This might be a by-product of the more peripheral location of the WA compared to the other fROIs, which being spatially closer to each other are facilitated to co-activate due to white fibers wiring them together. Wernicke's area is also the only area among the four ones in the common network which is not largely part of the maximum $k$-core (see Fig. \ref{fig:kcorehistogram}). This result is in agreement to what discussed above about this area and, again, might be due to the more perimetric location of the WA in the common network. 

Finally, we investigated the functional architecture of the BA anatomical sub-areas, revealing a different connectivity between tri-BA, op-BA and the other areas of the common network for this specific language task.
In Sec. \ref{sec:dis_common} we discuss our findings regarding the functional connectivity of the BA-subdivisions contextualizing them with known white matter connections that these areas share with the rest of the brain, found in other studies.

\subsection{Functional and structural connectivity of the common network}\label{sec:dis_common}

We observe that the left ventral preMA is the most connected area of the common network, with four total connections and  the strongest connectivity with op-BA ($W^C=0.74 \pm 31$) and with pre-SMA ($W^C=0.64 \pm 0.31$). As shown in Fig. \ref{fig:representative} for a representative subject, the ventral preMA is functionally connected to all the main cortical language areas of the dominant hemisphere, suggesting that this area may play an important role in speech production (other subjects show qualitatively the same feature, see Supplementary Fig. \ref{fig:20individuals}). 

%The contribution of the motor cortex to speech perception was studied by Wilson et al. \cite{wilson2004listening} using fMRI. The authors investigated motor activations in 10 healthy subjects {\color{red} listening to speech, pointing out a predominant activation of the preMA -- ASK LUCA}, with variable lateralization. 
Tate {\it et al}. \cite{tate2014probabilistic} investigated the crucial cortical epicenters of human language function by means of intraoperative direct cortical stimulation in 165 consecutive patients affected by low-grade glioma. The study shows that speech arrest is localized to the ventral preMA instead of the classical BA. Furthermore, the presence of gliomas growing in the left ventral preMA has been related to a higher percentage of speech deficits than gliomas infiltrating the classical BA, providing a possible clinical correlate of the results of Tate {\it et al}. \cite{tate2014probabilistic,bizzi2012aphasia}.  

However, one must be careful not to over interpret these results, as the highest connectivity does not necessarily imply a central or essential role of that particular fROIs in the network. Using advanced graph theoretical analysis, Morone {\it et al}. \cite{morone2015influence} demonstrated that the most connected nodes in a network often do not correspond to the most essential nodes, the elimination of which would lead to collapse of that particular network. This idea has been recently tested on functional networks obtained from fMRI of rodent brains and verified through {\it in-vivo} pharmaco-genetic intervention \cite{del2018finding}. 

Although the correspondence between structural and functional connectivity is not fully understood yet \cite{honey2009predicting}, the arrangement displayed by our study is supported by structural evidence. The existence of a physical connection between ventral preMA and BA seems realistic, given their spatial contiguity. Besides representing a shared origin for the main bundles of the dorsal pathway \cite{chang2015contemporary, dick2014language}, the two areas may be directly connected by a specific opercular-premotor fascicle (described in the next section) \cite{lemaire2013extended}.

The pre-SMA shows connectivity with both ventral preMA and BA (see Fig. \ref{fig:common3D}, Supplementary Fig. \ref{fig:20individuals} and Supplementary Table \ref{table:common}). These functional connections are consistent with the organization of the structural language connectome to some extent: the Frontal Aslant Tract (FAT), an association motor pathway that underlies verbal fluency and connects pre-SMA and BA \cite{catani2013novel,ford2010structural,jenabi2014probabilistic}, likely includes projection to posterior regions of the MFG, corresponding to the ventral preMA \cite{chang2015contemporary}.

The low connectivity weight between ventral preMA and WA (see Supplementary Table \ref{table:common}) may be explained by the increased distance between the two structures. Of note, we find that the functional connectivity weight between op-BA and WA is similar to that of the ventral preMA and WA (Supplementary Table \ref{table:common}), which is consistent with their structural connection through the same white matter tract, corresponding to the arcuate component of the AF/SLF system \cite{chang2015contemporary,dick2014language}.
	%Finally, the areas showing the lowest functional connectivity in our results are pre-SMA and WA (.........). Such finding is not surprising, since a direct structural connection has not been described between the two areas to date. 

\subsection{Broca's Area subdivisions}

Our findings show that the subdivisions of Broca's area present different patterns of connectivity within the language network, with the opercular portion appearing more connected to all the significant nodes of the common network compared to the triangular part. This evidence appears in line with the structural architecture of the network. 

The prominent interaction between ventral preMA and op-BA found in this study ($W^C=0.74 \pm 0.31$, see Supplementary Table \ref{table:common}) supports the evidence of a structural link between op-BA and preMA, as suggested by Lemaire {\it et al}.  using DTI analysis \cite{lemaire2013extended}. The authors investigated the structural connectome of the extended BA, identifying the U-shaped opercular-premotor fasciculus that connects the op-BA to the ipsilateral preMA \cite{lemaire2013extended}. On the contrary, tri-BA and ventral preMA showed lower functional connectivity ($W^C=0.37 \pm 0.29$), possibly suggesting indirect communication through the op-BA.

The second strongest functional connection between BA's subareas and other fROIs of the common network that we find is the link between op-BA - pre-SMA ($W^C=0.35 \pm 0.23$). These two areas are connected by the FAT \cite{ford2010structural}, which originates in the SMA/pre-SMA and terminates into the posterior-most aspect of the IFG \cite{catani2013novel}. Triangular BA and pre-SMA share a lower functional connectivity weight ($W^C=0.20 \pm 0.21$) compared to op-BA and pre-SMA, reflecting the anatomic boundaries of the FAT. 

Finally, the functional link between op-BA and WA is in line with the evidence of a dorsal pathway of language between op-BA and STG through the AF/SLF system (dorsal pathway II) \cite{friederici2011brain}.  \\

%In this case, the slightly difference in connectivity between op-BA - WA and tri-BA - WA (....... respectively), may reflect the different pathway that links the two: according to Federici et al., tri-BA is connected to the STG through the ECFS, which belongs to the ventral pathway (Friederici, 2011).

%clinical significance
%\section{Conclusions}

%The present study explores functional connectivity of language network nodes in healthy right-handed subjects by employing well-established statistical inference techniques applied to tb-fMRI data. The method described here allows us to understand the functional interdependence between language-related areas activated during the execution of a specific task, possibly differentiating between essential and secondary language areas, with promising applications in clinical practice.

%This analysis highlights functional connectivity among active cerebral areas in patients with different disorders from routinely acquired tb-fMRI sequences. In patients with brain tumors, this new method could complement the pre-surgical planning, providing additional information about organization and polarization of the individual language network to guide neurosurgical intervention. 
%\noindent
{\bf Data availability} \\

Data that support the findings of this study are publicly available and have been deposited in {\small \url{http://www-levich.engr.ccny.cuny.edu/webpage/hmakse/brain/}

\section{ACKNOWLEDGMENTS}

We thank Mehrnaz Jenabi for help with AFNI and FSL software and Medeleine Gene for help with mining the data. 

The funding support for this study was provided by the National Institute of Health (NIH), NIH-NIBIB R01 EB022720-01 (Makse and Holodny, PI's), NIH-NCI U54CA137788/U54CA132378, P30CA008748, the National Science Foundation (NSF) NSF-IIS 1515022 (Makse, PI), ISSNAF imaging chapter award 2018 (Pasquini, PI) and ESOR Bracco clinical fellowship 2018 (Pasquini, PI). 

}

\bibliographystyle{myunsrt}
\bibliography{HEALTHY}

\setcounter{section}{0}
\setcounter{figure}{0}

\renewcommand{\figurename}{{\bf Supplementary Figure}}
\renewcommand{\tablename}{{\bf Supplementary Table}}

\setcounter{table}{0}
\setcounter{figure}{0}
\renewcommand{\thetable}{S\arabic{table}}
\renewcommand\thefigure{S\arabic{figure}}
\renewcommand{\theHtable}{Supplement.\thetable}
\renewcommand{\theHfigure}{Supplement.\thefigure}

\clearpage
\onecolumngrid
\section*{\bf {\Large Supplementary Information}}

% FIGURE %%%%%%%%%%%%
\begin{figure*}[!h]
\includegraphics[width=0.8\textwidth]{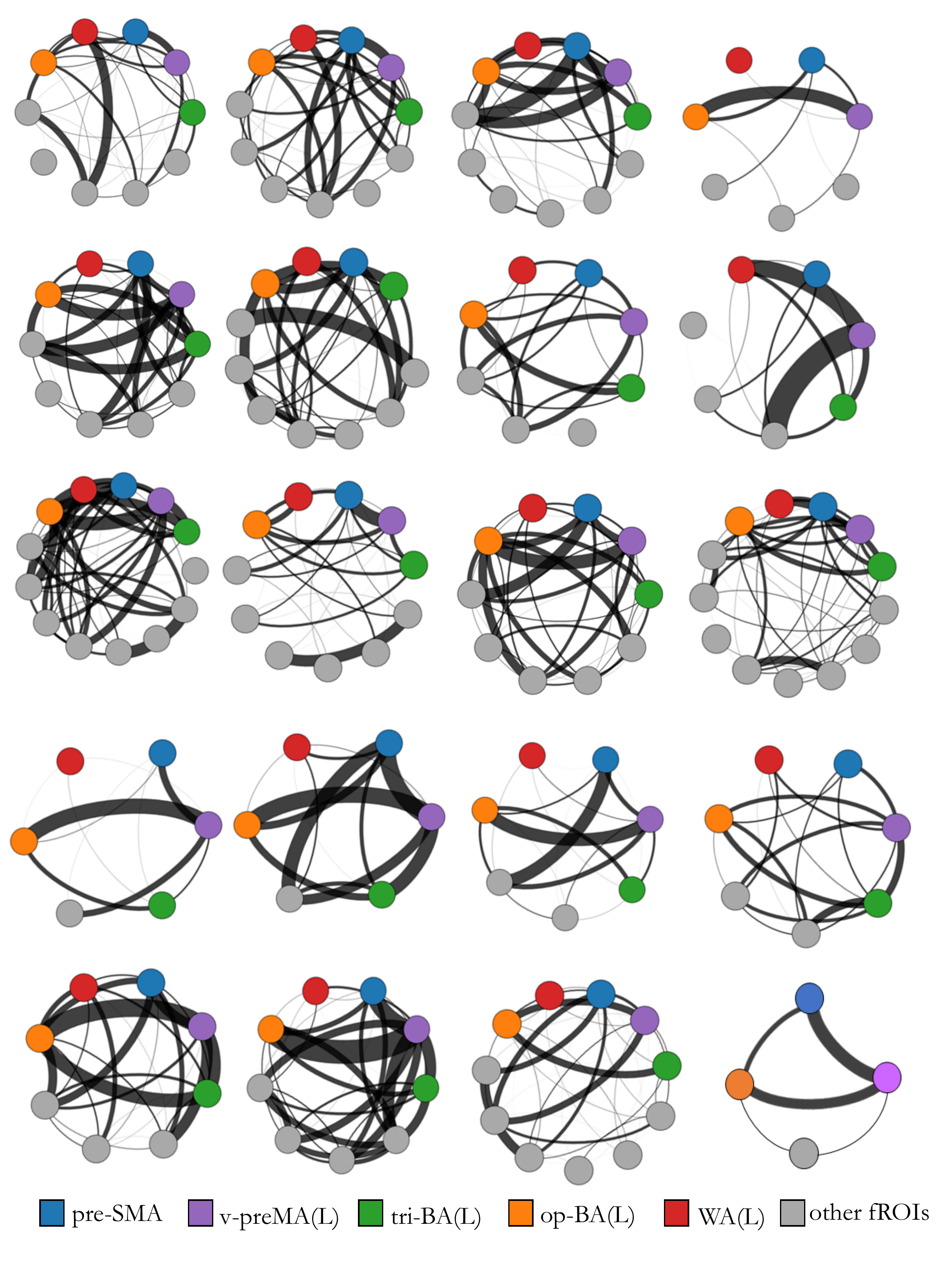}
\caption{{\bf Individual functional network for each subject.} Each node represents a fROIs in the individual network at the fROI-level, links are determined by Eq. \eqref{eq:W} and \eqref{eq:Wrescale} in the main text.
Functional ROIs (nodes) which are part of the common network (see Fig. \ref{fig:common3D}) are colored differently according to the legend, all other fMRI active areas are not distinguished and colored in grey. All the nodes (fROIs) are depicted with the same size  for illustration purpose and are not proportional to the actual fROI's size.
The construction of the functional network for each individual follows the procedure described in Sec. \ref{sec:individual} and in Fig. \ref{fig:representative}.}
\label{fig:20individuals}
\end{figure*}

\clearpage

\begin{table}[]
	\caption[Activated areas across subjects]{Activated areas across subjects.}
	\label{table:areas}
	\centering

	\begin{tabular}{|l|l|c|}
		\hline
		Activated area (fROI)             & Abbreviations & \multicolumn{1}{l|}{Activated in \# of subjects} \\ \hline
		Angular Gyrus (L)             & AngG (L)       & 4                                                \\ \hline
		opercularis-Broca's Area (L)  & op-BA (L)      & 19                                               \\ \hline
		triangularis-Broca's Area (L) & tri-BA (L)     & 18                                               \\ \hline
		Broca's Area (R)              & BA (R)         & 5                                                \\ \hline
		Caudate (L)                   & Caudate (L)    & 7                                                \\ \hline
		Caudate (R)                   & Caudate (R)    & 5                                                \\ \hline
		Deep Opercular Cortex (L)     & DOC (L)        & 3                                                \\ \hline
		Deep Opercular Cortex (R)     & DOC (R)        & 1                                                \\ \hline
		Middle Frontal Gyrus (L)      & MFG (L)        & 13                                               \\ \hline
		Middle Frontal Gyrus (R)      & MFG (R)        & 2                                                \\ \hline
		pre-Central Gyrus (L)         & pre-CG (L)     & 4                                                \\ \hline
		pre-Central Gyrus (R)         & pre-CG (R)     & 3                                                \\ \hline
		ventral-Premotor Area (L)   & v-preMA (L)    & 19                                               \\ \hline
		dorsal-Premotor Area (L)      & d-preMA (L)    & 10                                               \\ \hline
		Premotor Area (R)             & preMA (R)      & 2                                                \\ \hline
		pre-Supplementary Motor Area & pre-SMA       & 20                                               \\ \hline
		Supra-Marginal Gyrus (L)      & SupMG (L)      & 12                                               \\ \hline
		Supra-Marginal Gyrus (R)      & SupMG (R)      & 3                                                \\ \hline
		Wernicke's Area (L)           & WA (L)         & 19                                               \\ \hline
		Wernicke's Area (R)           & WA (R)         & 8                                                \\ \hline
	\end{tabular}
\end{table}

\begin{table}[]
	\caption[Links' weight between pairs of fROIs in the common network]{Links weight between pairs of fROIs in the common network ($W^C$)}
	\label{table:common}
	\centering

		\begin{tabular}{|c|c|}
			\hline
			Pair of connected fROIs & Link weights (mean $\pm$ stdv) \\ \hline
			v-preMA(L) - pre-SMA       & $0.64 \pm 0.31$            \\ \hline
			v-preMA(L) - WA(L)          & $0.18 \pm 0.24$            \\ \hline
			v-preMA(L) - op-BA(L)       & $0.74 \pm 0.31$            \\ \hline
			v-preMA(L) - tri-BA(L)      & $0.37 \pm 0.29$            \\ \hline
			pre-SMA - tri-BA(L)         & $0.20 \pm 0.21$            \\ \hline
			pre-SMA - op-BA(L)          & $0.35 \pm 0.23$            \\ \hline
			WA(L) - op-BA(L)            & $0.17 \pm 0.22$            \\ \hline
			op-BA(L) - tri-BA(L)        & $0.55 \pm 0.28$            \\ \hline
		\end{tabular}
	
\end{table}

\begin{center}
\begin{table*}[] 
\begin{center}
	\caption[Link's weight between pairs of fROI in each individual network]{Link's weight between pairs of fROI in each individual network ($\tilde{W}$)}
	\label{table:20individuals}
	\centering

\begin{tabular}{|l|c|c|c|c|c|c|c|c|}
			\hline
			\begin{tabular}[c]{@{}l@{}}fROI pairs\#/\\Subject  \end{tabular} & \multicolumn{1}{l|}{\begin{tabular}[c]{@{}l@{}}v-preMA(L) \\ -pre-SMA\end{tabular}} & \multicolumn{1}{l|}{\begin{tabular}[c]{@{}l@{}}v-preMA(L)\\ -WA(L)\end{tabular}} & \multicolumn{1}{l|}{\begin{tabular}[c]{@{}l@{}}v-preMA(L)\\ -op-BA(L)\end{tabular}} & \multicolumn{1}{l|}{\begin{tabular}[c]{@{}l@{}}v-preMA(L)\\ -tri-BA(L)\end{tabular}} & \multicolumn{1}{l|}{\begin{tabular}[c]{@{}l@{}}pre-SMA\\ -tri-BA(L)\end{tabular}} & \multicolumn{1}{l|}{\begin{tabular}[c]{@{}l@{}}pre-SMA\\ -op-BA(L)\end{tabular}} & \multicolumn{1}{l|}{\begin{tabular}[c]{@{}l@{}}WA(L)\\ -op-BA(L)\end{tabular}} & \multicolumn{1}{l|}{\begin{tabular}[c]{@{}l@{}}op-BA(L)\\ -tri-BA(L)\end{tabular}} \\ \hline
			1                                                                  & 0.46                                                                                & 0.01                                                                             & 1                                                                                   & 0.68                                                                                 & 0.65                                                                              & 0.1                                                                              & 0.24                                                                           & 0.65                                                                               \\ \hline
			2                                                                  & 0                                                                                   & 0                                                                                & 0                                                                                   & 0                                                                                    & 0.19                                                                              & 0.94                                                                             & 0.57                                                                           & 1                                                                                  \\ \hline
			3                                                                  & 1                                                                                   & 0.14                                                                             & 0.53                                                                                & 0.04                                                                                 & 0.14                                                                              & 0.61                                                                             & 0                                                                              & 0.44                                                                               \\ \hline
			4                                                                  & 0.43                                                                                & 0.16                                                                             & 1                                                                                   & 0.67                                                                                 & 0.14                                                                              & 0.73                                                                             & 0.79                                                                           & 0.71                                                                               \\ \hline
			5                                                                  & 1                                                                                   & 0.38                                                                             & 0.36                                                                                & 0.08                                                                                 & 0.28                                                                              & 0.1                                                                              & 0.2                                                                            & 0.16                                                                               \\ \hline
			6                                                                  & 0.55                                                                                & 0.31                                                                             & 0.33                                                                                & 0.18                                                                                 & 0.08                                                                              & 0.45                                                                             & 0.21                                                                           & 1                                                                                  \\ \hline
			7                                                                  & 0.19                                                                                & 0.02                                                                             & 1                                                                                   & 0                                                                                    & 0.00                                                                              & 0.3                                                                              & 0.02                                                                           & 0                                                                                  \\ \hline
			8                                                                  & 0.01                                                                                & 1                                                                                & 0                                                                                   & 0.41                                                                                 & 0.03                                                                              & 0                                                                                & 0                                                                              & 0                                                                                  \\ \hline
			9                                                                  & 1                                                                                   & 0.01                                                                             & 0.04                                                                                & 0.13                                                                                 & 0.62                                                                              & 0.19                                                                             & 0.09                                                                           & 0.45                                                                               \\ \hline
			10                                                                 & 1                                                                                   & 0                                                                                & 0.75                                                                                & 0                                                                                    & 0.00                                                                              & 0.38                                                                             & 0                                                                              & 0                                                                                  \\ \hline
			11                                                                 & 0.54                                                                                & 0.35                                                                             & 0.58                                                                                & 1                                                                                    & 0.23                                                                              & 0.21                                                                             & 0.01                                                                           & 0.76                                                                               \\ \hline
			12                                                                 & 0.58                                                                                & 0.03                                                                             & 1                                                                                   & 0.64                                                                                 & 0.07                                                                              & 0.35                                                                             & 0.27                                                                           & 0.75                                                                               \\ \hline
			13                                                                 & 1                                                                                   & 0.34                                                                             & 0.42                                                                                & 0.42                                                                                 & 0.58                                                                              & 0.33                                                                             & 0.07                                                                           & 0.66                                                                               \\ \hline
			14                                                                 & 0.41                                                                                & 0.06                                                                             & 0.97                                                                                & 0.1                                                                                  & 0.06                                                                              & 0.19                                                                             & 0.06                                                                           & 1                                                                                  \\ \hline
			15                                                                 & 0.47                                                                                & 0.01                                                                             & 1                                                                                   & 0.11                                                                                 & 0.01                                                                              & 0.04                                                                             & 0                                                                              & 0.29                                                                               \\ \hline
			16                                                                 & 0.34                                                                                & 0.05                                                                             & 1                                                                                   & 0.18                                                                                 & 0.00                                                                              & 0.13                                                                             & 0.01                                                                           & 0.33                                                                               \\ \hline
			17                                                                 & 1                                                                                   & 0.03                                                                             & 0.97                                                                                & 0.82                                                                                 & 0.23                                                                              & 0.45                                                                             & 0.02                                                                           & 0.42                                                                               \\ \hline
			18                                                                 & 0.58                                                                                & 0.15                                                                             & 1                                                                                   & 0.04                                                                                 & 0.01                                                                              & 0.57                                                                             & 0.09                                                                           & 0.35                                                                               \\ \hline
			19                                                                 & 0.7                                                                                 & 0.1                                                                              & 1                                                                                   & 0.5                                                                                  & 0.03                                                                              & 0.22                                                                             & 0.01                                                                           & 0.21                                                                               \\ \hline
			20                                                                 & 1                                                                                   & 0.03                                                                             & 0.36                                                                                & 0.33                                                                                 & 0.19                                                                              & 0.28                                                                             & 0.43                                                                           & 0.23                                                                               \\ \hline
		\end{tabular}	
\end{center}
\end{table*}
\end{center}

\end{document}